\documentstyle[epsfig,floats,amssymb,aps,pra,twocolumn]{revtex}

\begin{document}

\draft

\title{Three-dimensional vortex dynamics in Bose-Einstein condensates}

\author{B.~M.~Caradoc-Davies, R.~J.~Ballagh, and P.~B.~Blakie}
\address{Department~of~Physics, University~of~Otago, P.~O.~Box~56, Dunedin,
         New~Zealand.}

\date{March 6, 2000}

\wideabs{
\maketitle
\begin{abstract}
We simulate in the mean-field limit the effects of rotationally stirring a
three-dimensional trapped Bose-Einstein condensate with a Gaussian laser beam.
A single vortex cycling regime is found for a range of trap geometries, and is
well described as coherent cycling between the ground and the first excited
vortex states. The critical angular speed of stirring for vortex formation is
quantitatively predicted by a simple model. We report preliminary results for
the collisions of vortex lines, in which sections may be exchanged. 
\end{abstract}
}

The formation and properties of vortices in Bose-Einstein condensates are
attracting intense current interest. A number of schemes for producing vortices
in trapped condensates have been proposed theoretically, and the first
experimental creation of a topological vortex in a two-component condensate has
been reported \cite{MatthewsPRLSep1999}, in agreement with a numerical model 
\cite{WilliamsNatureOct1999}. Simulations of flow of a condensate past a
spherical object have shown vortex ring solutions \cite{WinieckiELDec1999}, and
numerical studies of vortex formation in rotating traps have shown the
formation of vortex arrays \cite{FederPRAJan2000}. Very recently the first
report has been made of the observation of vortices in a single-component
condensate \cite{MadisonPRLJan2000}.

\begin{figure}[t]
\begin{center}
\epsfbox{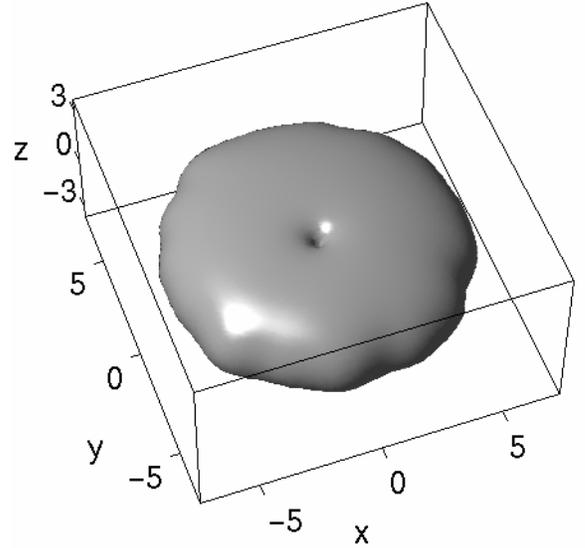}
\end{center}
\caption{Probability density isosurface ($|\psi|^2=10^{-4}$) for a condensate
stirred as described in the text, with the stirrer linearly withdrawn between
$t=7\pi$ and $t=8\pi$. Here $t=94.12$, and the hole near the centre of the
condensate is a vortex that has been drawn in from the edge and remains close
to to the centre of the condensate. Parameters are $C=1000$,
$\lambda=\sqrt{8}$, $W_0=4$, $w_s=4$, and $\rho_s=2$. The angular frequency of
stirring $\omega_f=0.3$ is less than the critical frequency $\omega_c=0.323$.}
\label{isocentral}
\end{figure}

In an earlier paper \cite{CaradocDaviesPRLAug1999}, we showed that a
rotationally stirred Bose-Einstein condensate, simulated in two dimensions,
exhibits a simple single vortex cycling regime, a behaviour we described as
nonlinear Rabi cycling. In the present paper we demonstrate that the single
vortex cycling regime is also found in full three-dimensional simulations of
condensate stirring (as shown in Fig.~\ref{isocentral}), in a parameter regime
accessible to current experiments. The single vortex regime is studied in a
variety of trap geometries (oblate to prolate) as is the dynamical stability of
central vortices. The line character of a vortex in three dimensions also gives
rise to some qualitatively new behaviour that has no analogue in two
dimensions. We show that in a spherical trap, stirring of the condensate can
produce two vortex lines which collide and exchange sections. 

As in the earlier work, our treatment is based on the time-dependent
Gross-Pitaevskii (GP) equation,
\begin{equation}
i\frac{\partial\psi({\bf r},t)}{\partial t} = \left[ -\nabla^2 + V({\bf r},t)
+ C|\psi({\bf r},t)|^2 \right] \psi({\bf r},t), \label{gpe}
\end{equation}
in dimensionless units \cite{Units}. The total external potential $V({\bf
r},t)$ is given by $(x^2+y^2+\lambda^2z^2)/4+W({\bf r},t)$, where the first
term represents a cylindrically symmetric trap with trap anisotropy parameter
$\lambda$. The second term $W({\bf r},t)$ is the contribution of the stirrer, a
far-blue detuned Gaussian laser beam cylindrically symmetric about a line
parallel to the $z$ axis,
\begin{eqnarray}
W({\bf r},t) = W_s(t)\,{\rm exp}\left[ -\left( 
  \frac{|\mbox{\boldmath$\rho$}-\mbox{\boldmath$\rho$}_{s}(t)|}{w_{s}/2}
\right)^2 \right] ,
\label{stirpot}
\end{eqnarray}
where $\mbox{\boldmath$\rho$}$ is the projection of ${\bf r}$ into the $z=0$
plane, and $\mbox{\boldmath$\rho$}_s$ is the centre of the Gaussian stirrer in
that plane.

\begin{figure}
\begin{center}
\epsfbox{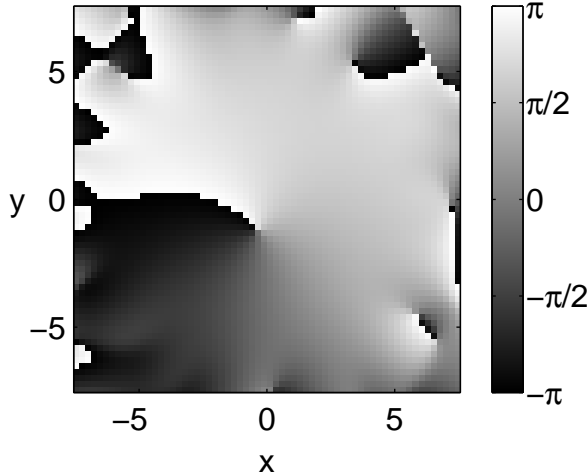}
\end{center}
\caption{Phase of $\psi$ in the $z=0$ plane for the condensate shown in
Fig.~\ref{isocentral}.}
\label{phasecentral}
\end{figure}

\begin{figure}
\begin{center}
\epsfbox{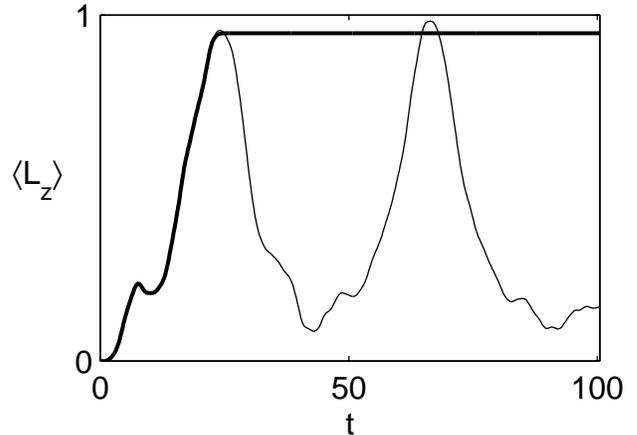}
\end{center}
\caption{Condensate angular momentum expectation value versus time, for the
case of a constantly rotating stirrer (thin line), or stirrer linearly
withdrawn between $t=7\pi$ and $t=8\pi$ (thick line). Parameters as in
Fig.~\ref{isocentral}.}
\label{angular}
\end{figure}

The condensate is stirred in a manner similar to our earlier work 
\cite{CaradocDaviesPRLAug1999}, with the stirrer moving in a circle of radius
$\rho_s$ at a constant angular velocity $\omega_f$. Here, however, we begin
with the ground state of the harmonic trap (with no stirrer present), and
minimise transient effects by increasing the stirrer amplitude $W_s$ linearly
from zero to a final value of $W_0$ between $t=0$ and $t=\pi$. With an
appropriate choice of parameters, as in Fig.~\ref{isocentral}, a single vortex
with positive circulation is drawn close to the centre of the condensate, and
for continuous stirring cycles in and out. This is exactly the behaviour found
previously in a two-dimensional system \cite{CaradocDaviesPRLAug1999}, and in
fact the parameter choice for the single vortex cycling behaviour in three
dimensions is guided by the same considerations outlined there, as we discuss
below. Figure~\ref{phasecentral}, where we plot the phase of the wavefunction,
shows that the central feature of Fig.~\ref{isocentral} is a vortex of positive
circulation. In Fig.~\ref{angular} the evolution of the angular momentum
expectation value (which in our dimensionless units is 0 for the ground state
and 1 for the central vortex state) is presented and confirms that for
continuous stirring (thin line), the vortex cycles in and out of the
condensate.

In contrast to the two-dimensional case, the three-dimensional simulation can
be unambiguously related to a realistic current experimental scenario. The
$C=1000$ case shown in Figs.~\ref{isocentral}--\ref{angular} corresponds to
$N=1.8\times 10^4$ atoms of $^{87}$Rb in the $|F=2,m_f=2\rangle$ hyperfine
state (for which we use an s-wave scattering length $a=5.29$~nm) in a time
orbiting potential (TOP) trap with radial trap frequency
$\omega_r=2\pi\times15$~Hz and $\lambda=\omega_z/\omega_r=\sqrt{8}$. For this
example, our computational units of time and length \cite{Units} correspond to
$10.6$~ms and $1.97$~$\mu$m respectively, and the Thomas-Fermi diameter of the
condensate in the radial direction of 11.7 corresponds to about 23~$\mu$m. The
stirrer Gaussian $1/e$ diameter $w_s=4$ corresponds to a Gaussian $1/e^2$ beam
diameter of approximately 11~$\mu$m. The required intensity of the stirring
laser is detuning dependent, but can be easily calculated from the light shift
potential $\hbar\Omega^2/4\Delta$, where $\Omega$ and $\Delta$ are the Rabi
frequency and detuning respectively of the atom-laser interaction. For example,
if the laser beam is 50~nm blue-detuned from the 780~nm atomic transition,
$\Delta=1.65\times10^{14}$~s$^{-1}$, and a laser power of 300~nW is required
for the 3~nK height stirrer we describe. The first angular momentum peak in
Fig.~\ref{angular} occurs at 0.26~s, which is much less than the lifetime of
the condensate. The occurrence of the single vortex cycling behaviour in this
experimentally accessible parameter regime suggests a method of preparing a
central vortex state; stir the condensate as we have described, then remove the
stirrer at the time when the angular momentum peak is expected. We illustrate
the implementation of this method in Fig.~\ref{angular} with the thick curve,
for which the stirrer is linearly withdrawn between $t=7\pi$ and $t=8\pi$. Once
the stirrer is removed, the vortex remains in the condensate, circling about
and close to the centre, and $\langle L_z \rangle$ is constant.

The vortex cycling behaviour illustrated by the example in
Figs.~\ref{isocentral}--\ref{angular} indicates that the two-state model
described in our earlier work (for the two-dimensional case
\cite{CaradocDaviesPRLAug1999}) also has application in three-dimensions. Here
we briefly outline the modifications required to extend the two-state model 
into three dimensions. We represent the condensate by an axisymmetric component
$\phi_s$, and a vortex-like component $\phi_v e^{i\theta}$ which has a
circulation about the $z$ axis, where $\phi_s$ and $\phi_v$ are real and
nonnegative, so
\begin{equation}
\psi({\bf r},t) = a_s(t)\phi_s(\rho,z,n_v) + 
                  a_v(t)\phi_v(\rho,z,n_v)e^{i\theta},
\label{ansatz}
\end{equation}
with $\rho=\sqrt{x^2+y^2}$. We choose $\phi_s$ and $\phi_v$ to satisfy the
following time-independent coupled equations,
\begin{mathletters}
\label{coupled}
\begin{eqnarray}
\mu_s\phi_s &=& \left[ 
-\frac{1}{\rho}\frac{d}{d\rho}\left(\rho\frac{d}{d\rho}\right) 
-\frac{d^2}{dz^2} \right. \nonumber \\
&& \left. \mbox{} + \frac{\rho^2+\lambda^2z^2}{4} +
C\left( n_s\phi_s^2 + 2n_v\phi_v^2 \right) \right]
\phi_s , \label{coupledsymmetric} \\ 
\mu_v\phi_v &=& \left[ 
-\frac{1}{\rho}\frac{d}{d\rho}\left(\rho\frac{d}{d\rho}\right) 
-\frac{d^2}{dz^2} + \frac{1}{\rho^2} \right. \nonumber \\
&& \left. \mbox{} + \frac{\rho^2+\lambda^2z^2}{4} +
C\left( n_v\phi_v^2 + 2n_s\phi_s^2 \right) \right]
\phi_v , \label{coupledvortex}
\end{eqnarray}
\end{mathletters}
where the eigenvalues $\mu_s$ and $\mu_v$ are found by solving 
Eqs.~(\ref{coupledvortex}) for a given choice of the vortex fraction
$n_v=|a_v|^2$. The quantity $n_s=|a_s|^2=1-n_v$. We note that when $n_v=0$,
Eqs.~(\ref{coupled}) reduce to Eq.~(\ref{coupledsymmetric}), the
time-independent GP equation for the ground state, while if $n_v=1$,
Eqs.~(\ref{coupled}) reduce to Eq.~(\ref{coupledvortex}), the time-independent
GP equation for the first excited vortex. The resulting nonlinear Rabi
equations for the two-state system [obtained by substituting Eq.~(\ref{ansatz})
into Eq.~(\ref{gpe})] are then identical in form to Eqs.~(5--6) of
\cite{CaradocDaviesPRLAug1999}, but with the wavefunctions $\phi_s$ and
$\phi_v$ now defined over $\rho$ and $z$ instead of $r$. 

The underlying assumption of our two-state model is that the ground state (or
more generally, $\phi_s$) is coupled primarily to the vortex state of lowest
energy with axis parallel to the stirrer axis. In two dimensions, there are a
limited number of excited condensate states in the appropriate energy range
that can be coupled to the symmetric state by the stirring potential, and we
have previously shown \cite{CaradocDaviesPRLAug1999} that the two state model
works well within a specified validity range. In three-dimensional
cylindrically symmetric traps, many more condensate eigenstates exist, and
although symmetry considerations limit the possible coupling from the symmetric
state, a two state model places more constraint on the description of the
system than in the two-dimensional case. We do not aim therefore to provide a
detailed representation of the stirring behaviour with our two state model,
instead we use it primarily to provide conceptual understanding of the cycling
behaviour. Nevertheless we also find it gives a quantitatively accurate
prediction of the critical frequency of rotation $\omega_c$, the stirring
frequency which causes a single vortex to cycle right to the centre of the
condensate. We have examined the single vortex cycling behaviour using the full
GP solution, for a variety of trap geometries including the oblate
($\lambda=\sqrt{8}$), spherical ($\lambda=1$), and prolate ($\lambda=1/3$)
cases, with a condensate of $C=1000$, and a stirring potential of the same size
and position as in Fig.~\ref{isocentral}. We find that the critical rotational
frequency for single vortex cycling is accurately predicted by the two state
model as the difference in energy between the ground and vortex states of the
condensate (modified by the finite size of the stirrer, see Eq.~(7) of
\cite{CaradocDaviesPRLAug1999}). As in \cite{CaradocDaviesPRLAug1999}, a
sufficiently large Rabi frequency is required to distort the energy barrier
between ground and vortex state enough to permit cycling. When the critical
frequency of rotation is exceeded, additional vortices penetrate the
condensate.

It is worth noting that as the $z$ component of the trapping potential become
weaker (that is, as $\lambda$ becomes smaller), then in addition to the vortex
state, the stirring potential increasingly excites the dipole (centre of mass)
oscillation of the condensate. This can be understood in terms of the Ehrenfest
theorem: for our choice of parameters the mean force $\langle -\nabla W\rangle$
of the stirrer increases as the condensate radial size decreases because the
condensate is increasingly concentrated near the steepest part of the stirring
potential. The centre of mass motion contributes angular momentum $\langle L_z
\rangle_{COM}$ to the condensate angular momentum, where
\begin{equation}
\langle L_z \rangle_{COM} = \langle x \rangle \langle P_y \rangle 
                          - \langle y \rangle \langle P_x \rangle .
\end{equation}
We find that $\langle L_z \rangle - \langle L_z \rangle_{COM}$ (which is a
measure of the angular momentum caused by the presence of the vortex) for all
three trap geometries cycles between 0 and 1 when $\omega_f$ is slightly less
than $\omega_c$ (the single vortex cycling regime) and oscillates about 1 when 
$\omega_f$ just exceeds $\omega_c$ (and multiple vortices penetrate the 
condensate). $\langle L_z \rangle_{COM}$ is small for the $\lambda=\sqrt{8}$ 
case shown in Fig.~\ref{angular}.

If the single stirrer used in Fig.~\ref{isocentral} is replaced by two stirrers
of half the height but on opposite sides of the $z$ axis, no vortex cycling
occurs, and $\langle L_z\rangle$ exhibits only small oscillations of amplitude
less than $0.04$. This is easily understood from our two state model, because
the Rabi frequency $\Omega$ (see Eq.~(6b) of \cite{CaradocDaviesPRLAug1999}) is
zero for this stirrer configuration, as it is for any stirrer configuration of
even symmetry. It is worth pointing out that our scenario of coherent cycling
should be distinguished from vortex production by damping in a trap with a
symmetric rotating distortion, such as the recently reported experiment
\cite{MadisonPRLJan2000}, or numerical damping schemes such as imaginary-time
propagation \cite{FederPRAJan2000}.

Our three-dimensional simulations also allow us to test (in the mean-field
limit) the stability of pure vortex eigenstates. Using as an initial state a
central vortex (an eigenstate of the time-independent GP equation) for
$C=1000$, a perturbation was applied by linearly inserting and withdrawing
(between $t=0$ and $t=\pi$) a stirrer of the same size as in
Fig.~\ref{isocentral} but at a fixed position in the laboratory frame. We find
that for a range of trap geometries ($\lambda=\sqrt{8}$, $1$, or $1/3$), the 
$\langle L_z \rangle=1$ vortex is dynamically stable, and remains near the 
centre of the condensate. We can understand this result in terms of Rabi
cycling. The stationary stirrer ($\omega_f=0$) provides a potential which is
far-detuned from resonance ($\omega_f=\omega_c$), and thus there is a low
probability of transfer out of the initial (vortex) state. Indeed if the
stirrer remains fixed in the condensate after insertion and is not withdrawn,
we find that the vortex still remains near the centre of the condensate and
follows a small closed path displaced slightly away from the stirrer. By
contrast, for these geometries, the $\langle L_z \rangle=2$ central vortex
eigenstate immediately dissociates into two unit vortices when perturbed. We
note that these stability results refer to dynamical stability in the GP limit
(that is, $T=0$), as opposed to thermodynamic stability
\cite{RokhsarPRLSep1997,FetterEprintAug1998}.

\begin{figure}[t]
\begin{center}
\epsfbox{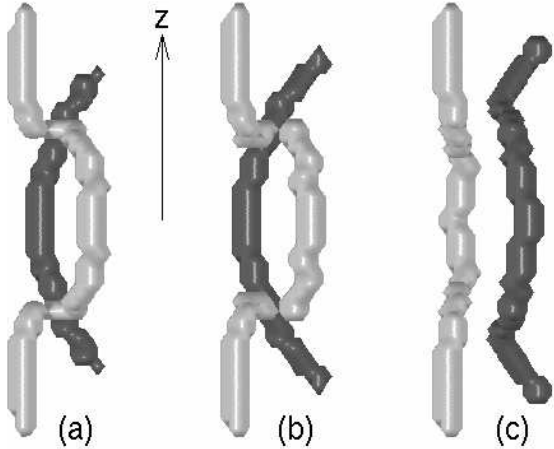}
\end{center}
\caption{Collision of vortex lines and exchange of sections in a stirred
condensate. View is from a position in the $z=0$ plane, looking almost down the
$y=-x$ line. In (a) $t=6.79$, (b) $t=6.91$, and (c) $t=7.04$. Parameters are
the same as in Fig.~\ref{isocentral} except that $\lambda=1$, The angular
frequency of stirring $\omega_f=0.45$ exceeds the critical frequency
$\omega_c=0.399$.}
\label{linecollision}
\end{figure}

In addition to confirming the basic features of the two state model of vortex
cycling, our three-dimensional simulations also reveal some new phenomena. The
vortices produced in three dimensions by stirring are line vortices. In the
$\lambda=\sqrt{8}$ configuration, the lines are nearly straight and parallel to
the stirrer axis, and when two lines of the same circulation form, we find they
repel each other. For a spherical trap ($\lambda=1$) however, vortex lines
along any direction are degenerate (in the absence of the stirrer) with the
vortex eigenstate about the $z$ axis. Accordingly, the vortex lines can become
more curved than in the oblate case, as we find in our simulations below the
critical frequency, where the vortex that forms can have appreciable curvature.
Above the critical frequency, when multiple vortices are drawn into the
condensate, there is a striking consequence of this potential for vortex lines
to deform. In a collision between two such curved vortex lines, sections of the
lines can be exchanged, as we show in the collision sequence in
Fig.~\ref{linecollision}. The condensate is stirred somewhat above the critical
frequency in a spherical trap, and by $t=6.79$ two vortex lines (both of
positive sense \cite{NotePositiveVortex}) have formed. The figure shows only
the vortex lines (detected numerically), which have been shaded light and dark
to distinguish one from the other. In Fig.~\ref{linecollision}(a) the dark
vortex line approaches the light vortex line from behind, and because of their
mutual repulsion, the light line has developed a bulge. In
Fig.~\ref{linecollision}(b) the dark line is now close to the light line,
causing the bulge to become recurved, with two small portions now nearly
antiparallel to the dark line. In Fig.~\ref{linecollision}(c) the two lines
have completed a collision and exchanged their central sections.

In conclusion, we have shown that the single vortex cycling regime we
discovered previously in simulations of two-dimensional condensates is also
present in three-dimensional condensates, for a range of trap geometries. We
have related our simulations to current experimental configurations, and have
demonstrated how the cycling behaviour could be used to generate a central
vortex state. We have investigated the dynamical stability of central vortex
states, and have also given initial evidence of the rich dynamical behaviour of
vortex lines.

This work was supported by the Marsden Fund of New Zealand under contract
PVT902.


\begin{thebibliography}{10}
\bibitem{Units}
Our units of time and length are $\omega^{-1}$ and $(\hbar/2m\omega)^{1/2}$
  respectively, where $\omega$ is the trap frequency in the $x$ direction. The
  nonlinearity parameter $C=8\pi a N (2m\omega/\hbar)^{1/2}$ and we normalise
  the wavefunction by $\int|\psi({\bf r},t)|^2d{\bf r}=1$.
\bibitem{NotePositiveVortex}
Positive sense here means that following from the bottom to the top the
  direction of circulation is given by the right hand rule.
\end{thebibliography}
\end{document}